\documentclass[%
 reprint,
superscriptaddress,
 amsmath,amssymb,
 aps,
nofootinbib
]{revtex4-2}

\usepackage{graphicx}
\usepackage[dvipsnames]{xcolor}
\usepackage{dcolumn}
\usepackage{bm}
\usepackage{hyperref}
\usepackage[normalem]{ulem}
\usepackage{makecell}
\usepackage{hhline}
\usepackage{caption} 
\newcommand{\SU}[1]{\ensuremath{\mathrm{SU}( #1 )}}

\newcommand{\SpR}[1]{\ensuremath{\mathrm{Sp}( #1,\mathbb{R} )}}
\newcommand{\hw}{\ensuremath{\hbar\Omega}}

\begin{document}

\preprint{APS/123-QED}
\title{Unexpected Rise in Nuclear Collectivity
from Short-Range Physics}

\author{Kevin S. Becker}
\affiliation{Department of Physics and Astronomy, Louisiana State University, Baton Rouge, LA 70803, USA}

\author{Kristina D. Launey}
\affiliation{Department of Physics and Astronomy, Louisiana State University, Baton Rouge, LA 70803, USA}

\author{Andreas Ekström}
\affiliation{Department of Physics, Chalmers Institute of Technology, Gothenburg, Sweden
}

\author{Tomáš Dytrych}
\affiliation{Department of Physics and Astronomy, Louisiana State University, Baton Rouge, LA 70803, USA}
\affiliation{Nuclear Physics Institute of the Czech Academy of Sciences, 250 68 \v{R}e\v{z}, Czech Republic}

\author{Daniel Langr}
\affiliation{Department of Computer Systems, Faculty of Information Technology, Czech Technical University in Prague, Prague 16000, Czech Republic}

\author{Grigor H. Sargsyan}
\affiliation{Facility for Rare Isotope Beams, Michigan State University, East Lansing, MI 48824, USA}

\author{Jerry P. Draayer}
\affiliation{Department of Physics and Astronomy, Louisiana State University, Baton Rouge, LA 70803, USA}

\date{\today}% It is always \today, today,

\begin{abstract}
We discover a surprising relation between the collective motion of nucleons within atomic nuclei, traditionally understood to be driven by long-range correlations, and short-range nucleon-nucleon interactions. Specifically, we find that quadrupole collectivity in low-lying states of $^6$Li and $^{12}$C, calculated with state-of-the-art \textit{ab initio} techniques, is significantly influenced by two opposing $S$-wave contact couplings that subtly alter the surface oscillations of one largely deformed nuclear shape, without changing that shape's overall contribution within the nucleus. The results offer new insights into the nature of emergent nuclear collectivity and its link to the underlying nucleon-nucleon interaction at short distances.
\end{abstract}

\maketitle

While the emergence of collective phenomena in atomic nuclei from elementary particle considerations has been recently shown in large-scale \textit{ab initio} calculations \cite{DytrychLDRWRBB20}, how exactly the underlying physics of quarks and gluons imparts collectivity remains elusive. Nuclear collectivity is historically understood to be driven by long-range correlations that deform the nucleus; deformed nuclei can rotate, which involves all particles in the system in a correlated motion \cite{BohrMottelson69}. Every nucleus can take on several shapes, typically one or two~\cite{DytrychLDRWRBB20}, and the deformation of each nuclear shape is often inferred from its equilibrium ``static'' deformation, whereas the ``snapshots'' of the shape's surface dynamics are called dynamical deformation (Fig.~\ref{fig:oscillations}). The extent of collectivity is informed by electric quadrupole moments $Q_2$, or equally, reduced transition strengths $B(E2)$, which increase with larger deformation \cite{Rowe_book16}. The prevalence of large deformation in most nuclei, even those with practically spherical ground states where shapes of different deformation co-exist \cite{HeydeW11}, is now evident from the vast body of experimental data \cite{GARRETT2022103931,Gray_2022,Ivanova_2024,Spataru_2024,Riczu_2024,Benjedi_2024,Gladnishki_2024,Kocheva_2024,Afanasjev_2024}. Furthermore, recent measurements of $B(E2)$ strengths have achieved such precision that they can now test microscopic approaches and the underlying inter-nucleon forces they adopt (see, e.g., \cite{Henderson:2017dqc, Ruotsalainen19}).

Yet, with no available nuclear force expressed directly in quark and gluon degrees of freedom, it remains unclear which parts of such fundamental interactions are responsible for the universal preference of large deformation, and are capable of substantially influencing collectivity. Fortunately, some insight emerges when the nuclear force is modeled with the state-of-the-art chiral effective field theory (EFT) (see, e.g., \cite{BedaqueVKolck02,EntemM03,Epelbaum:2014sza,RevModPhys.92.025004}). In particular, chiral EFT starts from nucleon and pion degrees of freedom, while accounting for the symmetry and symmetry-breaking patterns of the underlying theory of quantum chromodynamics (QCD). Chiral EFT hence provides a link to quark-gluon physics, which at low energies relevant to nuclei is averaged and encapsulated in its so-called low energy constants (LECs). The nuclear force thus partly depends on these unknown parameters which are typically fit to experimental few-nucleon data and might eventually be informed by QCD \cite{PhysRevC.97.021303,Tews_2020,PhysRevD.88.114507}. In this framework, we can directly probe the sensitivity of collective observables to the physics at very short distances (high energies) by studying the subset of LECs that determine the nucleon-nucleon (NN) contact interaction strengths.

\begin{center}
\captionsetup{type=figure}
  \includegraphics[width=0.75\linewidth]{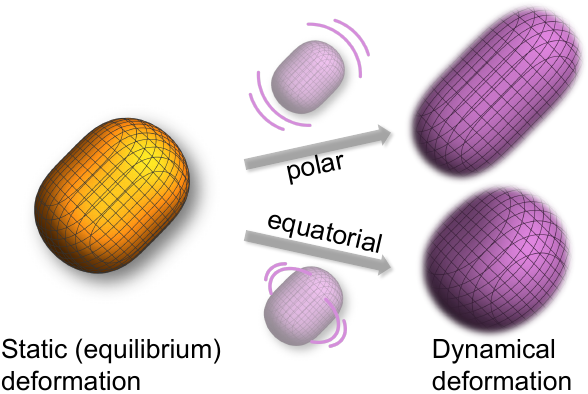}
  \caption{Schematic illustration, based on large-scale \textit{ab initio} calculations~\cite{DytrychLDRWRBB20}, of the static deformation (yellow) of a nuclear shape and two snapshots of its dynamical deformation (purple) arising from surface oscillations (particle-hole excitations): ``polar'' aligned with the symmetry axis, and ``equatorial'' in the orthogonal plane.
}
  \label{fig:oscillations}
\end{center}

In this letter, we report on a surprising new result: that nuclear collectivity can be enhanced not only through the traditional mechanism of shapes with ever-larger deformation, but also by a spatial redistribution of the surface oscillations of nuclear shapes that is mediated by the short-range $S$-wave NN contact interaction. Specifically, we discover that the LEC-independent forces, which include intermediate- and long-range interactions, determine the nuclear shapes and their contributions within a state, while at the same time variations of the short-range contact forces can further increase or decrease collectivity. The unexpected origin of this interesting effect, we find here, is subtle changes in the surface oscillations of one largely deformed and predominant shape.

We examine this by providing sensitivity and response analyses for nuclear energies and, for the first time, quadrupole moments, presented here for low-lying states of $^6$Li and $^{12}$C. Our results are made possible with the methodology of global sensitivity analysis (GSA) \cite{SOBOL2001271,SALTELLI2010259}, which is often prohibitively expensive since the number of model evaluations required for convergence grows rapidly with the number of model parameters. Fortunately, the symmetry-adapted no-core shell model (SA-NCSM)  \cite{LauneyMD_ARNPS21,LauneyDD16, DytrychSBDV_PRL07} alleviates the cost of each evaluation by utilizing a fraction of ultra-large no-core shell model spaces without compromising the accuracy of the results. The SA-NCSM is an \textit{ab initio} approach that has provided successful descriptions of electric quadrupole moments \cite{LauneyMD_ARNPS21} and transition strengths without utilizing effective charges \cite{Henderson:2017dqc, Ruotsalainen19,PhysRevC.100.014322}, energy spectra \cite{Sargsyan_A8,heller2022new}, clustering features \cite{Sargsyan_A8,DreyfussLESBDD20,DreyfussLTDBDB16}, and even reaction dynamics \cite{MercenneLEDP19,MercenneLDEQSD21,LauneyMD_ARNPS21,BurrowsSumRules}, up through the calcium region. It produces the same results as the traditional no-core shell model \cite{NavratilVB00,BarrettNV13} for a given inter-nucleon interaction, characterized by the number of harmonic oscillator (HO) shells accessible to the nucleons, and the HO inter-shell energy $\hbar \Omega$. 

Within this framework, we probe the response of collective features to short-distance NN coupling strengths, given by eleven LECs in chiral two-body potentials that are associated with the contact interactions of two nucleons for a given partial wave $^{2S+1}\ell_J$. These LECs are hence denoted as $C_{^{2S+1}\ell_{J}}$. Our new insights into the particular details of the nuclear force inform deformation and collectivity, and are therefore critical in advancing the frontier of high-precision nuclear simulations and uncertainty quantification that stem from the parametrizations of chiral potentials (see, e.g.,~\cite{PhysRevX.6.011019,PhysRevC.107.014001,becker2024uncertainty}).

\noindent
{\it Resilience of nuclear shapes.--} To study the responses of collective features to variations in the short-range coupling strengths, we start with the LEC parametrization of the NNLO$_{\rm opt}$ chiral NN potential~\cite{Ekstrom13}. This potential is used without three-nucleon forces, which have been shown to contribute minimally to three- and four-nucleon binding energies \cite{Ekstrom13}, and is parametrized by $14$ LECs from Feynman diagrams included up to next-to-next-to-leading order (NNLO) in the chiral expansion. Furthermore, NNLO$_{\rm opt}$ has been found to reproduce various observables and yield results equivalent to those obtained from chiral potentials that require three-nucleon forces, including, e.g., the $^4$He electric dipole polarizability \cite{BakerLBND20} and $A=8$ energy spectra and quadrupole moments \cite{Sargsyan_A8} (cf. Ref.~\cite{MarisVN13} for $A=7,8$ with the N$^3$LO-EM chiral potential~\cite{EntemM03} with a corresponding, comparatively significant, three-nucleon force that decreases $Q_2$ by up to $\sim 10\%$). 

An NNLO chiral NN potential includes the following LECs: 11 contact couplings, $C^{(\textrm{LO})}_{^1S_0(pp, np, nn),^3S_1}$ at leading order (LO), $C^{(\textrm{NLO})}_{^1S_0,^3S_1}$ and $C_{^3S_1-^3D_1,^3P_{0,1,2},^1P_1}$ at next-to-leading order (NLO), along with $c_{1,3,4}$ at NNLO which parametrize parts of the two-pion exchange interaction (Fig. \ref{sp3r_su3_JJ2_wfn}, inset). We note that NNLO$_{\textrm{opt}}$ does not include pure $D$-wave contact interactions as they emerge at the next chiral order (N3LO), and further note that by using NNLO$_{\rm opt}$ this study adopts its regulator function and cutoff, variations of which are beyond the present scope. We vary these LECs, uniformly sampled using a Latin hypercube design \cite{lhs}, with each LEC bounded within $\pm 10\%$ of its NNLO$_{\rm opt}$ value.  

\begin{center}
\captionsetup{type=figure}
  \includegraphics[width=\linewidth]{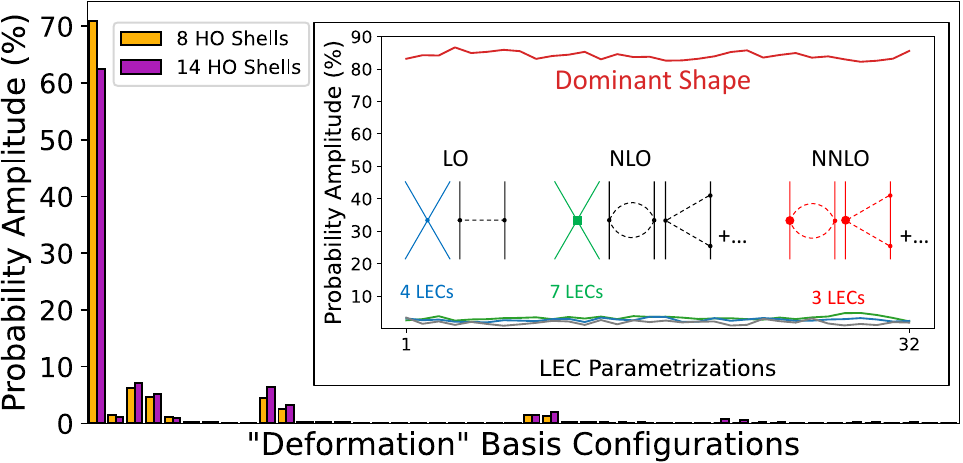}
  \caption{Probability amplitudes of the ``deformation'' basis configurations of the $^6$Li $1^+_{\rm g.s.}$ ground state calculated with NNLO$_{\rm opt}$ in complete model spaces of $8$ (yellow) and $14$ (purple) HO shells (shown are contributions with probability amplitude $\geq 0.1\%$). Inset: The four nuclear shapes with the largest probability amplitudes in the $8$-shell $^6$Li $1^+_{\rm g.s.}$ state across $32$ NN parametrizations uniformly sampled within $\pm10\%$ around the NNLO$_{\rm opt}$ LECs (see text for details). While not shown, the inset results remain practically unchanged for $1\%$-$50\%$ LEC variations \cite{becker_cgs2023,becker2024uncertainty}. Also shown are the Feynman diagrams for chiral NN forces up to NNLO with the total number of LECs per order.
  }
  \label{sp3r_su3_JJ2_wfn}
\end{center}

First, using NNLO$_{\rm opt}$, we perform SA-NCSM calculations --with no \emph{a priori} approximations or assumptions-- in the so-called ``deformation'' basis, known as the \SU{3}-adapted basis \cite{LauneyDD16,LauneyDSBD20}, for complete model spaces, that is, with all possible basis states given the number of accessible HO shells. Clearly, as shown in Fig. \ref{sp3r_su3_JJ2_wfn} for the $1^+_{\rm g.s.}$ ground state of $^6$Li, the probability amplitudes calculated in a model space of $8$ HO shells vary only slightly compared to those calculated in the drastically larger model space spanned by 14 HO shells. Similar outcomes are observed for the energy $E_{\rm X}$ of the first excited $3^+_1$ state of $^6$Li, as well as the quadrupole moments $Q_2$ of the $1^+_{\rm g.s.}$ and $3^+_1$ (see Fig. \ref{obs_convergence} in the supplemental material). These results indicate that the $8$-shell model spaces are sufficient for our GSA that requires hundreds of thousands of large-scale computations, especially since the focus here is on responses to variations in the LECs and the physics that tracks with such changes. In addition, without loss of generality, our calculations use a single $\hbar \Omega = 15$ MeV, for which the rms radii converge comparatively faster \cite{DytrychLDRWRBB20}.

\begin{figure*}
  \centering
  \includegraphics[width=\linewidth]{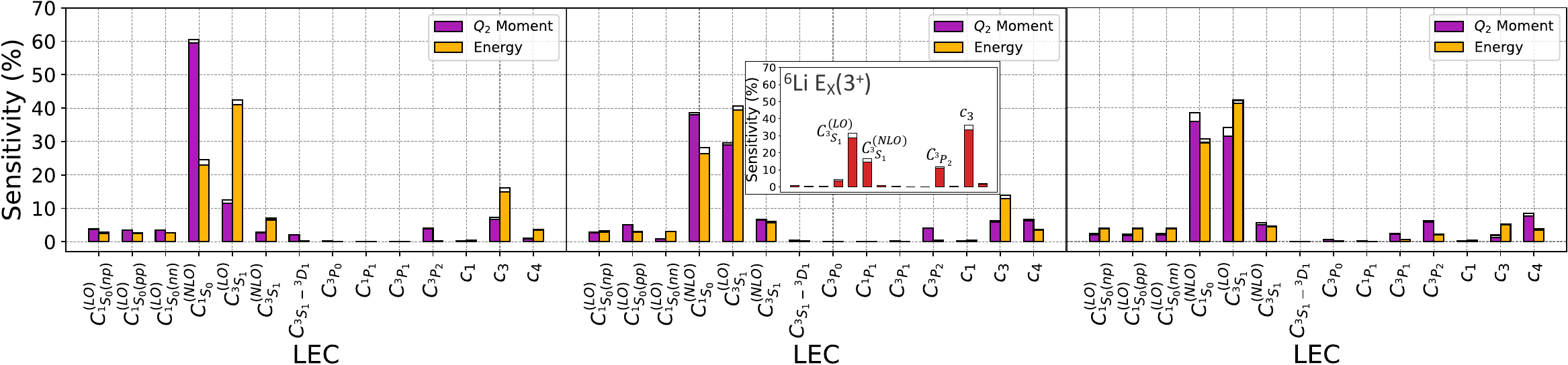}
\caption{First-order (colored) and total-order (white) sensitivity indices of the quadrupole moments (purple) and energies (yellow) of (a) the $^6$Li $1^+_{\rm g.s.}$, (b) $^6$Li $3^+_1$, and (c) $^{12}$C $2^+_1$ states. Calculations use SA model spaces comprised of fourteen shapes spanning $8$ HO shells. Inset: Sensitivity indices for the $^6$Li $3^+_1$ excitation energy, $E_X(3^+)$.
}
\label{GSA_results}
\end{figure*}

A very interesting result emerges when we further project these states onto the so-called ``shape'' basis, known as the \SpR{3}-adapted basis \cite{DytrychSBDV_PRL07,LauneyDD16,DytrychLDRWRBB20,LauneyMD_ARNPS21}. \SpR{3}-preserving subspaces represent nuclear shapes \cite{RosensteelR77,Rowe85,Rowe_book16}, each of which includes a static deformation and its dynamical surface oscillations [multiples of 2\hw~1-particle-1-hole (1p-1h) excitations], as illustrated in Fig.~\ref{fig:oscillations}. Remarkably, as shown in the inset of Fig. \ref{sp3r_su3_JJ2_wfn}, the shapes, and above all the dominant shape contributing $82.2$-$86.6\%$ to the $1^+_{\rm g.s.}$ state, are left practically untouched across the 10\% LEC variations (see Fig. \ref{shape_resilience} of the supplemental material for all the states under consideration). We note that for the NNLO$_{\rm opt}$ parametrization, the most dominant shape comprises about $86\%$ of the $^6$Li wave functions, whereas fourteen shapes recover about $97\%$ of the total probabilities. In a simple picture of rigid shapes that rotate (as is assumed in Ref.~\cite{andreas_r42_updated}), the quadrupole moment will only vary as these probabilities vary \cite{Rowe85,heller2022new}. Given the resilience of the nuclear shapes to LEC variations, should one then expect merely marginal effects on quadrupole moments? 

To answer this question, we first perform a global sensitivity analysis to probe whether $Q_2$ is sensitive to changes in the LECs individually, or if the induced variance depends strongly on correlations between the LECs. If the correlations are small, we can then search for the underlying mechanism through the responses of $Q_2$ to each LEC independently.

\noindent
{\it Sensitivity of collective observables.--} We generate $300,000$ samples in the LEC parameter space using Saltelli's procedure, implemented by the python library SALib \cite{Iwanaga2022, Herman2017}. With the aid of high-performance computing, these parametrizations are used for the state-of-the-art evaluation of 300,000 nuclear simulations -- only possible in the \SpR{3} shape basis -- of the energy and $Q_2$ moment of the $1^+_{\rm g.s.}$ and $3^+_1$ states of $^6$Li, and the $2^+_1$ state of $^{12}$C. We find a pronounced sensitivity of $Q_2$ to two singlet (spin-zero) and triplet (spin-one) $S$-wave contacts, namely, $C^{(\textrm{LO})}_{^3S_1}$ and $C^{(\textrm{NLO})}_{^1S_0}$ (Fig. \ref{GSA_results}). This is indicated by the first-order sensitivity indices $S_i$, quantifying the fractional variances in the samples due to the $i^{\textrm{th}}$ LEC alone. In contrast to the excitation energies, this is the same physics to which their binding energies are most sensitive, as earlier recognized in $^{16}$O \cite{Ekstrom19}, suggesting that both the binding energy and quadrupole moment may simultaneously inform the LEC fitting.
 
To enable this huge number of calculations, required for convergence of the GSA results, we utilize symmetry-adapted (SA) model spaces selected from fourteen shapes spanning $8$ HO shells. We find that the resulting sensitivity patterns are practically unchanged across various model space selections and sizes, and show this for $^6$Li in model spaces comprised of $1$ shape, $14$ shapes, and all possible shapes (corresponding to the complete model space), and in larger model spaces comprised of $10$ and $12$ HO shells in Figs.~\ref{GSA_results_suppl} and \ref{GSA_convergence} of the supplemental material, respectively. Based on this validation and using the same shape-selection prescription, the sensitivity analysis becomes feasible for the $^{12}$C $2^+_1$ state in SA model spaces consisting of fourteen shapes that account for about $80\%$ of the state. In this way, one can utilize Hamiltonian matrices in the selected shape basis of highly reduced dimensions (e.g., four orders of magnitude reduction for the $2^+_1$ state) that can serve as emulators, similarly to, e.g., Refs.~\cite{odell2023rose,Dj_rv_2022,beckerled22,konig20,PhysRevLett.121.032501,RevModPhys.96.031002}. Indeed, we find that the sensitivity patterns for $^{12}$C [Fig. \ref{GSA_results}(c)] closely resemble those for $^6$Li. 

Importantly, the total-order sensitivity indices, which additionally include correlations with all remaining LECs, practically coincide with the first-order indices $S_{i}$ (Fig.~\ref{GSA_results}). This suggests that the main effect of each LEC on $Q_2$ is largely decoupled from those of the remaining parameters, and can be explored separately. This is especially useful since the sensitivity indices are normalized to the total variance of the $Q_2$ distributions sampled, and thus provide a relative contribution only. The degree to which each contact coupling affects $Q_2$ can hence be understood by studying the responses of the LECs individually, as discussed next.

\begin{figure*}
  \centering
  \includegraphics[width=\linewidth]{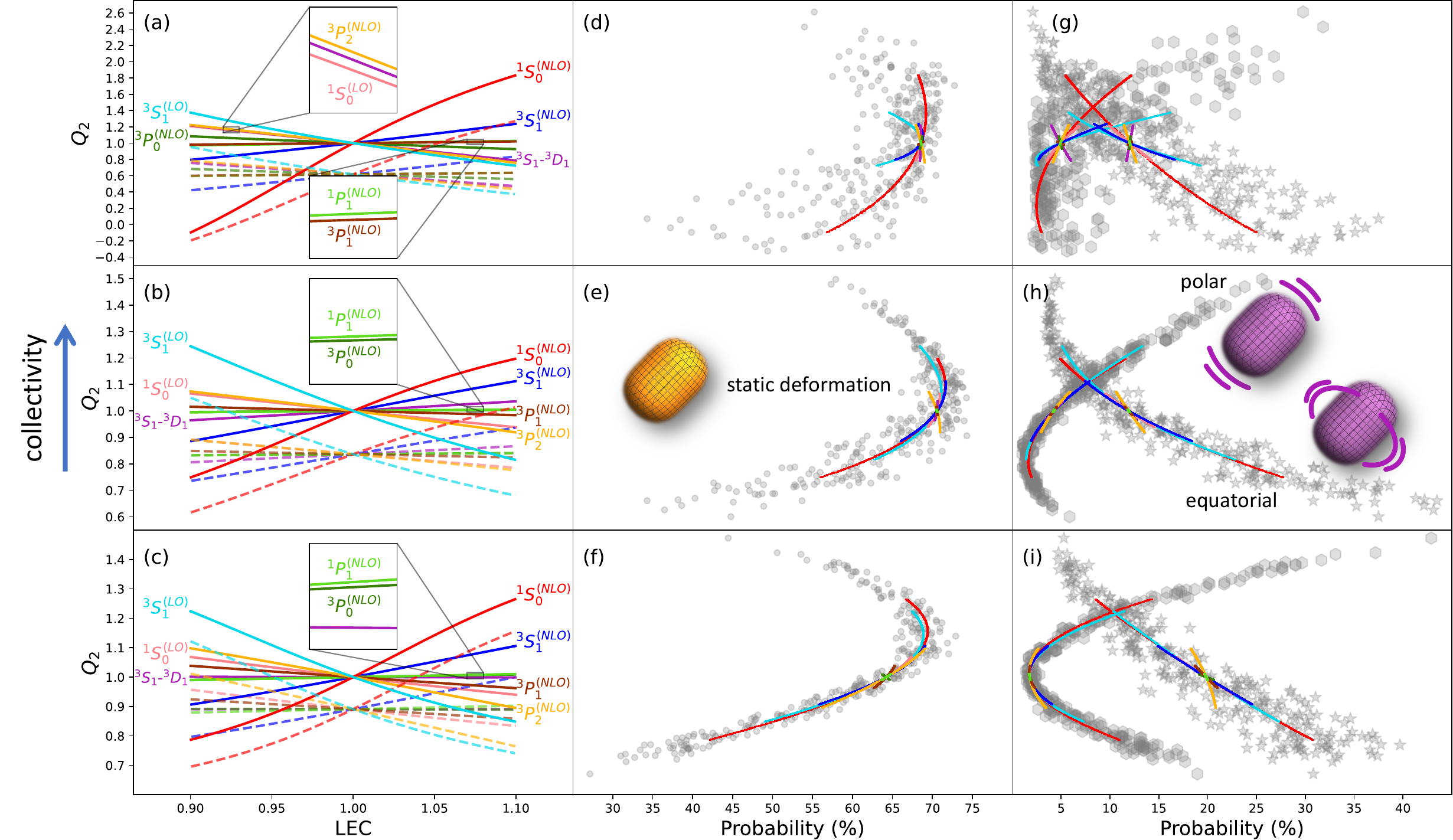}
\caption{
$Q_{2}$ quadrupole moment (solid), relative to their NNLO$_{\rm opt}$ values, of (a) the $^6$Li $1^+_{\textrm{g.s.}}$, (b) $^6$Li $3^+_1$, and (c) $^{12}$C $2^+_1$ states vs. each LEC varied $\pm 10\%$ around its NNLO$_{\rm opt}$ value. Also shown is the $Q_{2}$ contribution of the dominant shape (dashed). The corresponding probability amplitudes of the dominant shape are shown for (d)-(f) the static (0p-0h) deformation, together with (g)-(i) the surface oscillations along the symmetry axis (labeled ``polar'') and perpendicular to it (labeled ``equatorial''). $Q_2$ moments obtained from 300 simultaneously varied LEC samples are shown in gray for the static (circles), polar (hexagons), and equatorial deformation (stars).
}
\label{Q2_vs_LECs}
\end{figure*}

\noindent
{\it Short-range footprints on long-range physics.--} We discover that nonnegligible nuclear quadrupole moments of collective states, such as the $^6$Li $3^+_1$ and $^{12}$C $2^+_1$ states, can be enhanced by nearly 50\% through two competing $S$-wave contact interactions, which do so via an unexpected mechanism (Fig.~\ref{Q2_vs_LECs}). To show this, we first fix all LECs but one to their NNLO$_{\rm opt}$ values, and draw $300$ samples for each individual LEC, following a Latin hypercube design (Fig.~\ref{Q2_vs_LECs}, where for visual clarity we exclude the LO proton-proton and neutron-neutron $^1S_0$ contacts, and since we focus on the contact physics, similarly exclude $c_1$, $c_3$, and $c_4$; for completeness, we provide all these in the supplemental material).

Corroborating the GSA results of Fig. \ref{GSA_results}, Fig. \ref{Q2_vs_LECs} unveils additionally interesting features, namely that $Q_2$ depends practically linearly on the LECs, and that the S-wave contacts act in opposition: increasing the LO S-wave couplings (especially the triplet $^3S_1$) of the attractive delta contact interaction decreases $Q_2$, while increasing the coupling strength of the repulsive NLO S-wave contact (dominated by the singlet $^1S_0$) results in larger $Q_2$ moments. We note that the stronger the LO $^3S_1$ coupling, the more bound the deuteron system is, and that an attractive delta interaction has historically been used to describe pairing. Indeed, in phenomenological nuclear models, increasing the pairing strength leads to some decrease in $Q_2$, which actually results from a strong mixing of nuclear shapes \cite{ESCHER1998662}. Surprisingly, Fig.~\ref{Q2_vs_LECs} shows such a decrease in $Q_2$, but with practically no change in the mixing of the nuclear shapes, as illustrated in Fig.~\ref{sp3r_su3_JJ2_wfn}. This suggests that another mechanism might be responsible. Given the large probability amplitudes of the dominant nuclear shapes, we explore whether the observed variances are tied to changes within these dominant shapes. Importantly, since the total quadrupole moment is the sum of $Q_2$ contributions from each nuclear shape (as $Q_2$ does not couple different shapes), the component of $Q_2$ attributed to the dominant shape not only yields most of the total $Q_2$ value, but also exhibits nearly identical LEC dependence [Fig. \ref{Q2_vs_LECs}, (a)–(c), dashed].

The physics of the dominant shape unveils, for the first time, the unexpected result that the S-wave contacts of the underlying chiral potential can significantly alter the magnitude of $Q_2$ by impacting the shape's surface dynamics: the magnitude of $Q_2$ increases for a stronger repulsive $^1S_0$ contact interaction by favoring polar oscillations, that is, multiples of two HO quanta along the symmetry axis, and decreases as the equatorial modes (having HO quanta in the plane perpendicular to the symmetry axis) become dominant with a more attractive $^3S_1$ contact [Fig. \ref{Q2_vs_LECs}(g)-(i)]. This can be understood as follows: the $^1S_0$ contact, by spreading nucleons to higher HO shells, reduces Pauli principle constraints and makes prolate (football-like) configurations accessible, and these are known to have low kinetic energy \cite{CASTEL1990121, ROWE1970273} (Fig.~\ref{fig:expecval}). In contrast to this, as the LO $^3S_1$ coupling of the delta contact increases, excitations in the equatorial planes become favorable, leading to a further decrease of the negative potential energy (Fig.~\ref{fig:expecval}); and while this results in larger binding energy, it comes at the expense of reduced collectivity. 

These two contacts together can impact $Q_2$ by up to $\sim 50\%$ for the $^6$Li $3^+_1$ and $^{12}$C $2^+_1$ collective states (see Fig.~\ref{fig:expecval} for the $3^+_1$ state), compared to, e.g., $2$-$15\%$ many-body model uncertainties in the present GSA calculations (cf. Refs.~\cite{shin_2017,Grigorthesis} for the infinite-space estimates using NNLO$_{\rm opt}$ for $^6$Li and $^{12}$C; in addition, Ref.~\cite{Grigorthesis} reports $13\%$ many-body model uncertainties in the \textit{ab initio} estimate for $Q_2$). Furthermore, the importance of the polar modes cannot be understated: whether the equilibrium deformation is prolate as in $^6$Li, or oblate as in $^{12}$C, only polar oscillations will enhance collectivity. In short, we link the impact of the short-distance contact physics on nuclear collectivity to the surface dynamics that can increase (decrease) $Q_2$ predominantly through a stronger repulsive NLO (attractive LO) S-wave contact that favors polar (equatorial) oscillations, which lower the kinetic (potential) energy. This impact is not merely marginal, which in turn makes quadrupole moments important observables to constrain realistic nuclear forces. 

\begin{center}
\captionsetup{type=figure}
  \includegraphics[width=\linewidth]{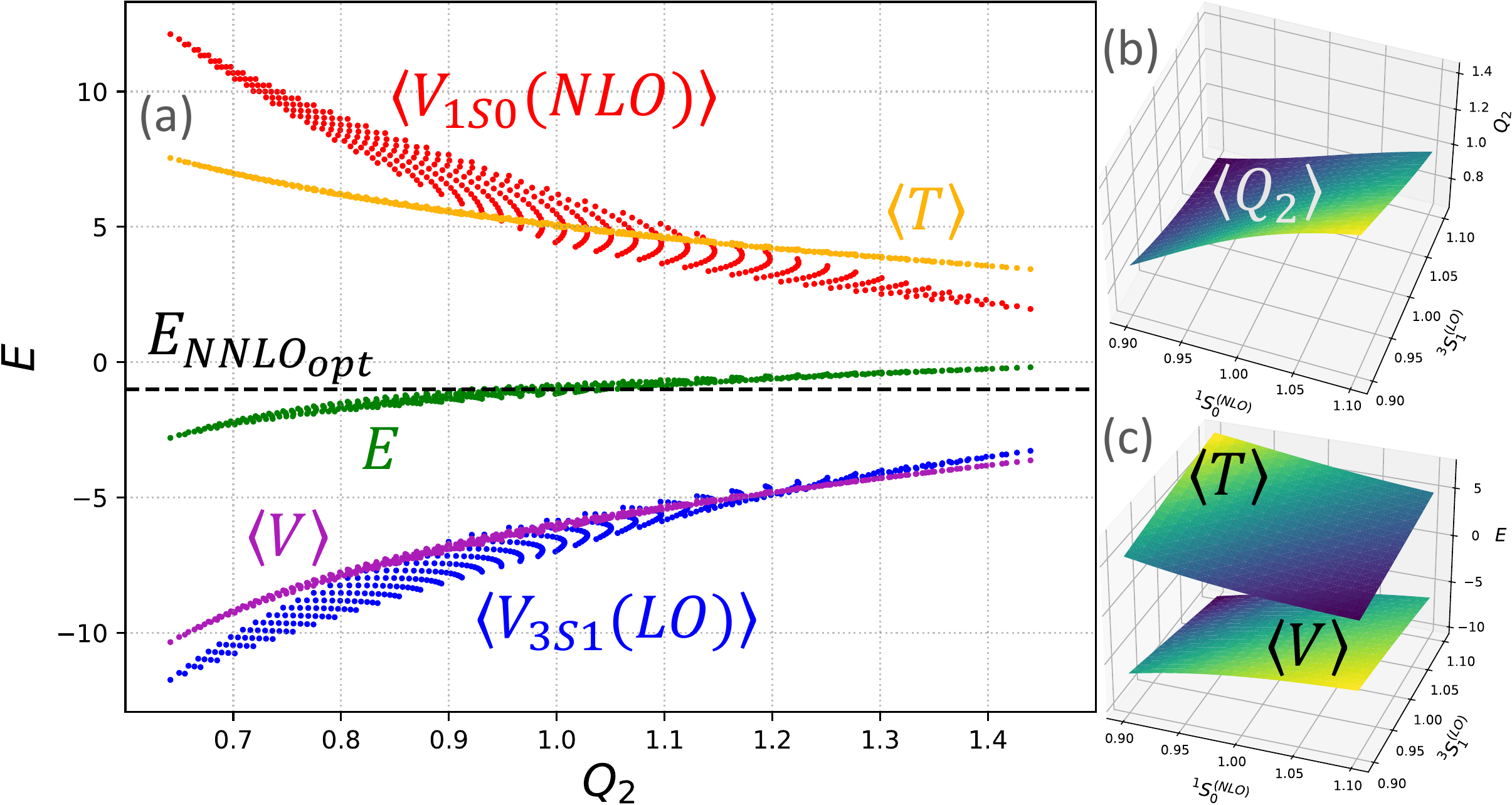}
  \caption{(a) The energy $E$ of the $^6$Li $3^+_1$ state (green), the expectation value of the total potential energy $\langle V \rangle$ (purple) and total kinetic energy $\langle T \rangle$ (yellow), as well as of the LO $^3S_1$ and NLO $^1S_0$ potentials (blue and red, respectively), as functions of the electric quadrupole moment $Q_2$. The quadrupole moments are normalized to their NNLO$_{\mathrm{opt}}$ values, while the expectation values are all normalized to the NNLO$_{\mathrm{opt}}$ energy of the $3^+_1$ state, $E_{\rm NNLO_{\rm opt}}$. (b) The quadrupole moment and (c) $\langle T \rangle$ and $\langle V \rangle$ vs. the LO $^3S_1$ and NLO $^1S_0$ contact couplings.
  }
  \label{fig:expecval}
\end{center}

To summarize, we find that electric quadrupole moments of light deformed nuclei, which provide a measure of their collectivity, are sensitive to short-range chiral contact NN interactions, namely, two opposing attractive triplet and repulsive singlet S-wave contacts. While historically larger $Q_2$ moments have been associated with reduced mixing between nuclear shapes or shapes of larger deformation, here, for the first time, we show that contact interactions can significantly increase the magnitude of $Q_2$ by subtly altering the surface dynamics of (typically) one dominant nuclear shape. We show that quadrupole collectivity is maximized by enhancing the polar surface oscillations at the expense of the remaining excitations, and this is most readily accomplished by increasing the magnitude of the $^1S_0$ coupling or decreasing that of the $^3S_1$ contact. We have thus found a direct link between an \textit{ab initio} description of nuclear collectivity and the short-range part of the underlying strong forces between nucleons which are rooted in QCD. This, in turn, paves the way for the construction of chiral potentials that can predict collective properties of atomic nuclei, while at the same time shedding new light on the emergent collectivity that dominates nuclear dynamics.

\begin{acknowledgments}
This work was supported by the U.S. Department of Energy (DE-SC0023532), the European Research Council (ERC) under the European Union's Horizon 2020 research and innovation program (Grant Agreement No. 758027), the Swedish Research Council (Grant Agreement No. 2020- 05127), and the Czech Science Foundation (22-14497S). This material is based upon work supported by the U.S. Department of Energy, Office of Science, Office of Nuclear Physics, under the FRIB Theory Alliance award DE-SC0013617. This work benefited from high performance computational resources provided by LSU (www.hpc.lsu.edu), the National Energy Research Scientific Computing Center (NERSC), a U.S. Department of Energy Office of Science User Facility at Lawrence Berkeley National Laboratory operated under Contract No. DE-AC02-05CH11231, as well as the Frontera computing project at the Texas Advanced Computing Center, made possible by National Science Foundation award OAC-1818253.
\end{acknowledgments}

\bibliography{main_ARXIV}% 
%\newpage
\section{Supplemental Materials}

\subsection{Sobol' global sensitivity analysis (GSA)}

For completeness, we provide a brief overview of Sobol' sensitivity analysis (for further details, we refer the reader to the works of Sobol' and Saltelli, e.g., \cite{SOBOL2001271, SALTELLI2002280, SALTELLI2010259}). The basic premise is to determine how uncertainty in the inputs of a given model induce uncertainties in its output, at various orders of correlation. In order to fully explore these high-dimensional correlations, we employ a quasi-random sampling technique based on Sobol' sequences developed by A. Saltelli. This method generates a set of inputs that efficiently fills the entire hypervolume of the parameter space \textit{and} each lower-dimensional surface, according to the desired sample size ($300,000$ in our study) and the allowed ranges of each parameter. This allows for a comparatively fast calculation of the Sobol' sensitivity indices, which become increasingly more difficult to compute at higher orders, via Monte Carlo integration. This is especially useful for the so-called total-order effects or indices, which can be well approximated with a single integral rather than through explicit summation of all lower-order terms \cite{HOMMA19961}.

The Sobol' indices themselves are defined as the fractional variances of the sampled model outputs when considering variations on a subset of its inputs, achieved by decomposing the total sample variance $\mathcal{V}$ in the following way: 
\begin{equation}
    \mathcal{V} = \sum_i \mathcal{V}_i + \sum_{i,j} \mathcal{V}_{ij} + ... + \mathcal{V}_{ij...d},
\end{equation}
where $d=14$ is the dimension of the parameter space, $\mathcal{V}_i$ corresponds to the partial variance due to only the $i^{\textrm{th}}$ LEC, $\mathcal{V}_{ij}$ arises through simultaneous variations of the $i^{\textrm{th}}$ and $j^{\textrm{th}}$ LECs, etc. The Sobol' indices are computed by dividing these partial variances by the total variance $\mathcal{V}$, thus the first-order indices $S_i$ are given by $\mathcal{V}_i /\mathcal{V}$, the second-order indices by $\mathcal{V}_{ij}/\mathcal{V}$, and so on. The total-order effect for parameter $i$ is hence obtained by summing all fractional variances indexed by $i$, i.e., 
\begin{equation}
    S_{Ti} = S_i + \sum_j S_{ij} + \sum_{j,k} S_{ijk} + ... + S_{ij...d},
\end{equation}
though as mentioned this demanding sum is approximated with a single Monte Carlo integration (for further technical details, we refer to \cite{SALTELLI2002280, HOMMA19961}). Sobol' analysis is nonetheless extremely computationally intensive for large parameter spaces, as is the case in this work, and the need to lower the cost of individual model evaluations without sacrificing accuracy quickly becomes evident.
\begin{center}
  \includegraphics[width=0.8\linewidth]{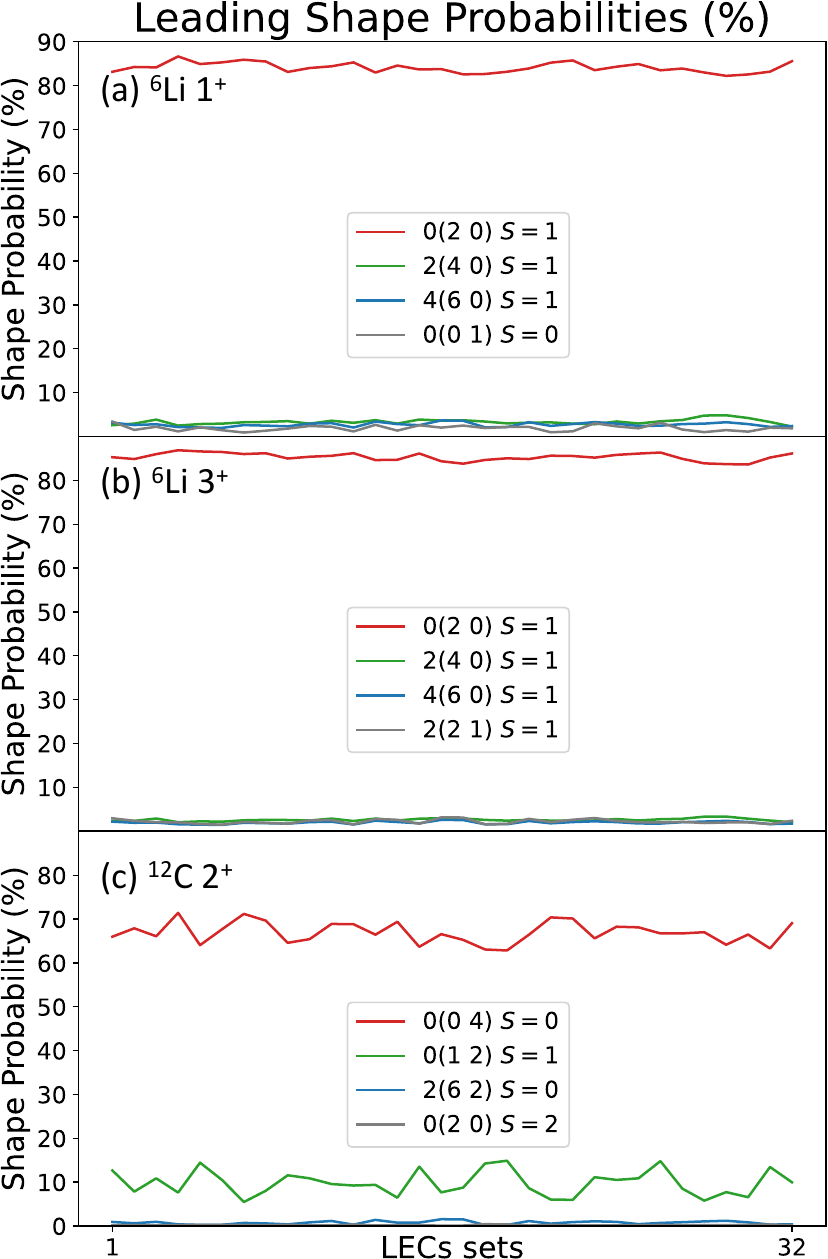}
\captionsetup{type=figure}
  \caption{Probabilities of the four most dominant shapes in (a) the $^6$Li $1^{+}_{\textrm{g.s.}}$ ground state, (b) its first excited $3^{+}_1$ state, and (c) the first excited $2^{+}_1$ state of $^{12}$C, across the same $32$ parametrizations shown in the inset of Fig. \ref{sp3r_su3_JJ2_wfn}. The wavefunctions for all three states are calculated with $\hbar \Omega = 15$ MeV and complete 8-shell HO model spaces. Each shape is labeled by the quantum numbers of its \SpR{3} irreducible representation: $N_{\sigma}(\lambda \, \mu)_{\sigma} \,S$, where $N_{\sigma}$ is the total number of HO quanta for the static deformation (with $N_{\sigma}=0$ denoting the valence shell), $(\lambda \, \mu)_{\sigma}$ indicates the \SU{3} quantum numbers describing the static deformation, and $S$ is the total intrinsic spin.
}
\label{shape_resilience}
\end{center}

% 6Li, Nmax 6 sensitivities vs SA selection
\begin{figure*}
  \centering
  \includegraphics[width=\linewidth]{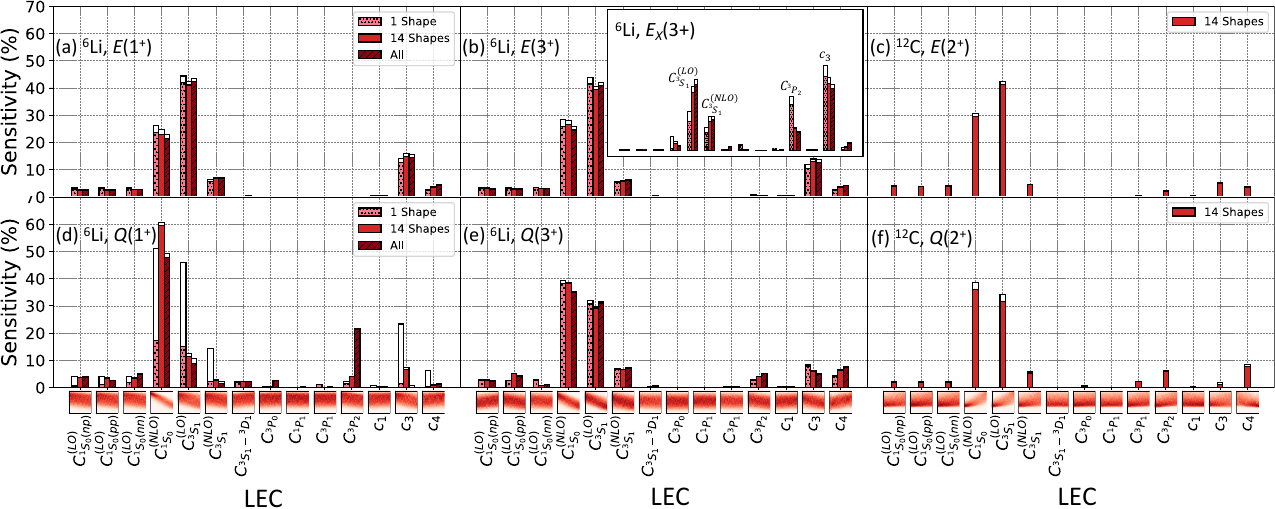}
\caption{
First-order (colored) and total-order (white) sensitivity indices of the energies (top panels) and quadrupole moments (bottom panels) of the (a) \& (d) $^6$Li $1^+_{\rm g.s.}$, (b) \& (e) $^6$Li $3^+_1$, and (c) \& (f) $^{12}$C $2^+_1$ states. Calculations use SA model spaces comprised of 1 shape (light red, dotted; labeled as ``1 Shape''), fourteen shapes (red, solid; ``14 Shapes''), and all possible shapes (dark red, slashed; ``All'', which is the complete no-core shell-model space) spanning $8$ HO shells. Histograms of the sampled LEC values and quadrupole moments divided into equally-spaced bins are shown at the bottom. Inset: Sensitivity indices for the $^6$Li excitation energy, $E_X$, for the $3^+_1$ state. 
}
\label{GSA_results_suppl}
\end{figure*}
\subsection{Dominant shapes in chiral potential parametrizations}

By organizing conventional no-core shell model spaces into symmetry-preserving subspaces of definite shape (``shape" basis configurations), the SA-NCSM reveals nature's striking preference for symmetry in light- and medium-mass nuclei \cite{DytrychLDRWRBB20,LauneyMD_ARNPS21}. This enables accurate calculations of nuclear observables using mere fractions of ultra-large conventional model spaces through the procedure of shape selection. Furthermore, since the approach utilizes laboratory particle coordinates, the ``shape" basis allows for an exact factorization of the center-of-mass motion, and states contaminated by spurious center-of-mass excitations are easily removed \cite{Verhaar60,Hecht71}.

For completeness, we demonstrate in Fig.~\ref{shape_resilience} the resiliency of the most dominant nuclear shapes in the $3^+_1$ and $2^+_1$ states, and compare to the results shown for the $1^+_{\textrm{g.s.}}$ in Fig. \ref{sp3r_su3_JJ2_wfn} in the main article. The same set of $32$ uniformly-sampled LECs sets are used for each calculation. It is clear that symplectic symmetry is a very good symmetry in all three states, as it is always the same shape that dominates even if its probability fluctuates. The results for the $3^+_1$ are extremely similar to those of the $1^+_{\textrm{g.s.}}$, with a prolate shape contributing $\geq 80\%$ for all LEC samples, and the next most-dominant shapes contribute $< 10\%$ each. In $^{12}$C, it is an oblate shape that dominates at approximately a $60\%\sim70\%$ level, while the second most impactful shape is near-oblate and contributes between approximately $5\%\sim 15\%$. The remaining two shapes are barely visible, and their probabilities fluctuate less appreciably than the first two. 

\subsection{Convergence of sensitivity indices with shape selection}

To determine the impact of shape selection on the sensitivity results, we perform the same analysis of the $^6$Li states in two additional SA model spaces consisting of all possible shapes (i.e., the complete no-core shell model space) and the single most dominant shape alone (respectively labeled ``All'' and ``1 shape'' in Fig. \ref{GSA_results_suppl}), both expanded up to $8$ HO shells. Clearly, Fig. \ref{GSA_results_suppl} shows that the complete model spaces yield very similar sensitivity indices to those calculated with the fourteen shapes discussed in the text and reported in Fig. \ref{GSA_results}. In addition, in all cases of Fig. \ref{GSA_results_suppl}(a), (b), \& (e), the sensitivity pattern remains the same even in the case of a single (dominant) shape. This is highly significant, strongly hinting that the observed variances are tied to changes in the deformed configurations belonging to that dominant shape, as demonstrated in Fig. \ref{Q2_vs_LECs}.

We note that the $Q_2(1^+_{\rm g.s.})$ moment for $^6$Li [Fig. \ref{GSA_results_suppl}(d)] exhibits some dependence on the model-space selection, e.g., the complete space yields different variances for $C_{^3P_2}$ and $c_3$. Further, the large differences between the first- and total-order indices when considering only the dominant shape suggest correlations between the LECs, likely because of the limited number of available configurations. These differences may stem from the fact that the wave function is predominantly $L = 0$ with small mixing from $L=2$ configurations ($\sim 10\%$). Since the $Q_2$ operator cannot couple two states with $L = 0$, its expectation value is close to zero and highly attuned to the extent and composition of the $L=2$ admixture~\cite{DytrychLMCDVL_PRL12}, suggesting greater sensitivity to this mixing. This results in a $Q_2(1^+_{\rm g.s.})$ that is two orders of magnitude smaller compared to the $^{6}$Li $3^+_1$ and $^{12}$C $2^+_1$ collective states. 

\subsection{Convergence with model space size}

To ensure that our results do not depend strongly on the number of included HO shells, we extend the 14-shape sensitivity analysis of the quadrupole moment for the $^6$Li $3^+_1$ state to $10$ shells, and the $1^+_{\rm{g.s.}}$ state to $12$ shells, that is, to larger model space sizes (Fig. \ref{GSA_convergence}). Clearly, the first- and total-order sensitivity indices for the quadrupole moments of both states are well converged.

\begin{center}
\captionsetup{type=figure}
  \includegraphics[width=\linewidth]{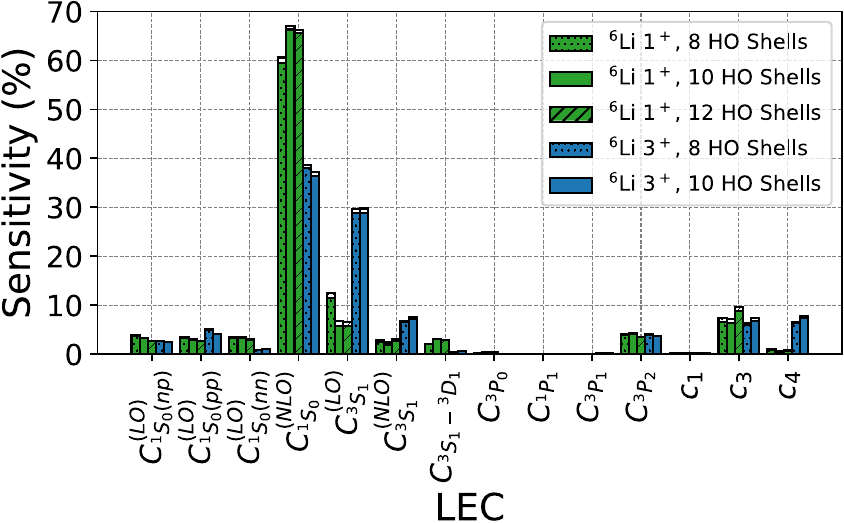}
  \caption{First-order (colored) and total-order (white) sensitivity indices of the quadrupole moments sampled in the $^6$Li $1^+_{\textrm{g.s.}}$ (green) and first excited $3^+_1$ states (blue). The shading patterns correspond to the 14-shape SA-model space expanded up to $8$ (dotted), $10$ (solid), and $12$ (slashed) HO shells.}
  \label{GSA_convergence}
\end{center}

\begin{center}
  \includegraphics[width=0.9\linewidth]{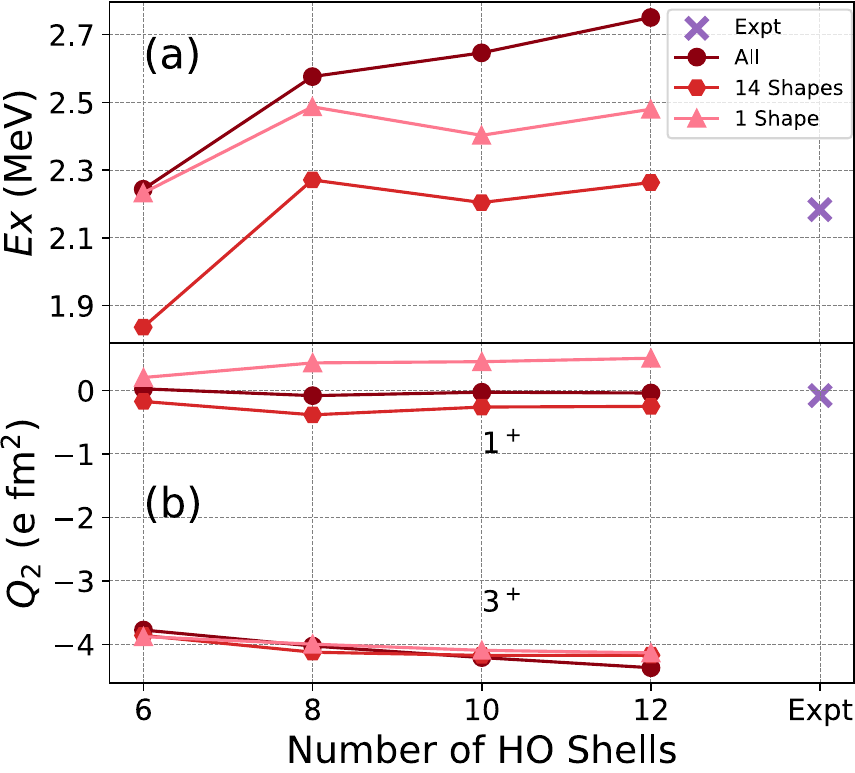}
\captionsetup{type=figure}
  \caption{(a) Excitation energy of the $^6$Li $3^{+}_1$ state as a function of the number of included HO shells, in SA spaces spanned by the most dominant shape (pink triangles), fourteen shapes (red hexagons), and all shapes (crimson circles). (b) The quadrupole moment of the $1^+_{\textrm{g.s.}}$ (top) and $3^+_1$ (bottom) states as a function of the number of included HO shells. All calculations use the NNLO$_{\rm opt}$ LEC parametrization and $\hbar \Omega = 15$ MeV. Experimental data is shown with purple crosses where available.
}
\label{obs_convergence}
\end{center}

\begin{figure*}
  \centering
  \includegraphics[width=\linewidth]{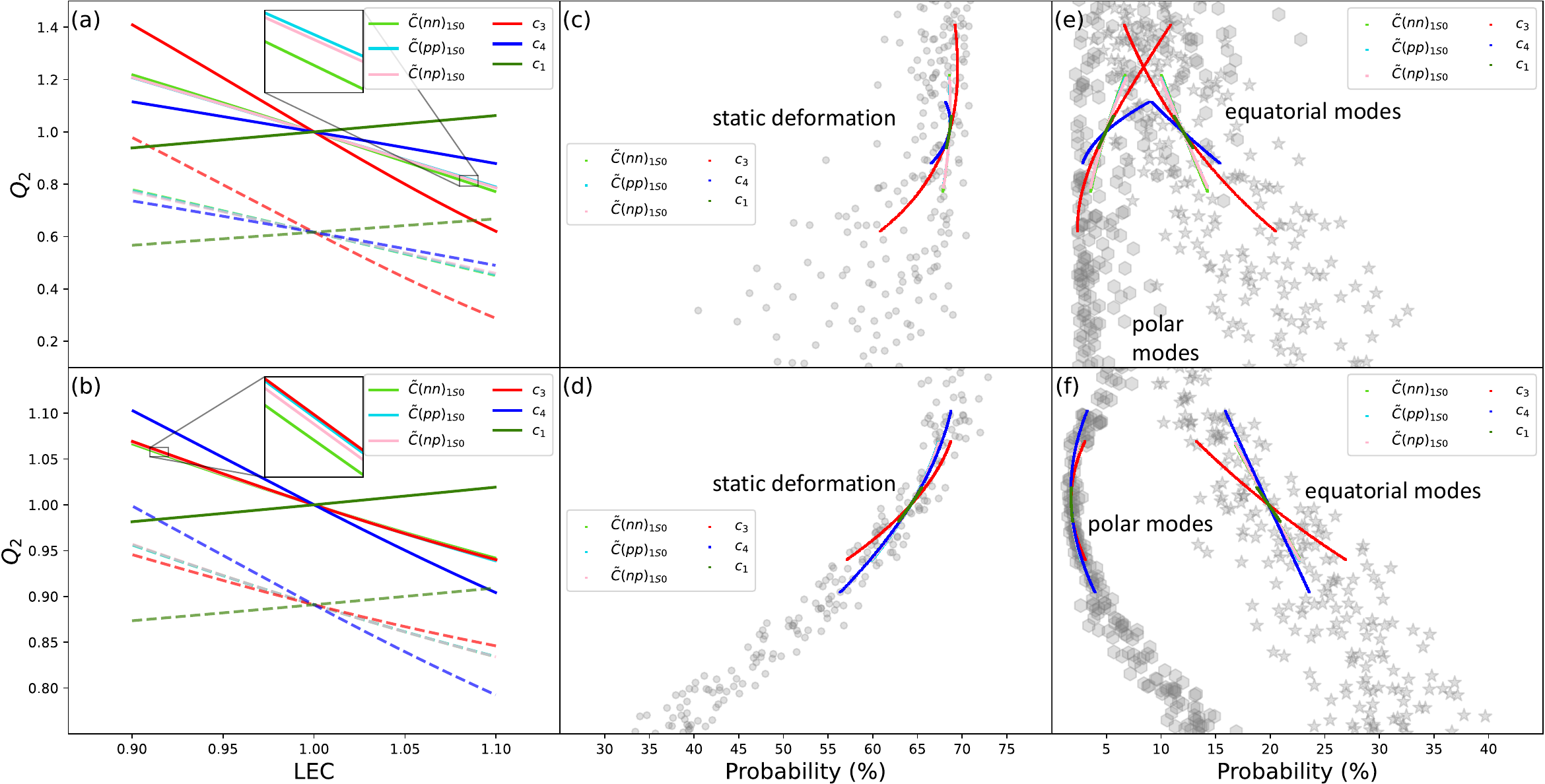}
\caption{$Q_{2}$ quadrupole moments (solid), relative to their NNLO$_{\rm opt}$ values, of (a) the $^6$Li $1^+_{\textrm{g.s.}}$, (b) $^6$Li $3^+_1$, and (c) $^{12}$C $2^+_1$ states vs. each LEC varied $\pm 10\%$ around its NNLO$_{\rm opt}$ value. Also shown is $Q_{2}$ of the dominant shape (dashed). The corresponding probability amplitudes of the dominant shapes are shown for (d)-(f) the static (0p-0h) deformation, together with (g)-(i) the surface oscillations along the symmetry axis (labeled ``polar'') and perpendicular to it (labeled ``equatorial''). $Q_2$ moments obtained from 300 simultaneously varied LEC samples are shown in gray for the static deformation (circles), polar (hexagons), and equatorial modes (stars).
}
\label{Q2_vs_LECs_supp}
\end{figure*}

In addition, we show that the $^6$Li $3^+_1$ excitation energy along with the quadrupole moment for the $3^+_1$ and $1^+_{\textrm{g.s.}}$ states, calculated using NNLO$_{\rm opt}$, exhibit converging trends with respect to the number of HO shells included, for all shape selections considered (Fig. \ref{obs_convergence}).

\vspace{8pt}
For the sake of completeness, we additionally show in Fig. \ref{Q2_vs_LECs_supp} how the total and single-shape $Q_2$ moments of the $^6$Li $1^+_{\textrm{g.s.}}$ and $^{12}$C $2^+_1$ states depend on the LECs excluded from Fig. \ref{Q2_vs_LECs}: the proton-proton and neutron-neutron components of $C^{(LO)}_{^1S_0}$, and the two-pion exchange LECs $c_1$, $c_3$, and $c_4$ (while not shown, the outcome is similar for the $3^+_1$ state of $^6$Li). We also include the proton-neutron component of $C^{(LO)}_{^1S_0}$ previously shown in Fig. \ref{Q2_vs_LECs}, to illustrate the minimal effect of isospin-invariance breaking in the leading order $^1S_0$ interaction. This set of LECs shown in Fig.~\ref{Q2_vs_LECs_supp} exerts only a marginal effect on both sets of quadrupole moments, especially compared to the dominant NLO $^1S_0$ and LO $^3S_1$ terms discussed in the main article (we note that the scale of the y-axis in Fig. \ref{Q2_vs_LECs_supp} is smaller than the one in Fig. \ref{Q2_vs_LECs}). 

\end{document}